# NUMERICAL SIMULATION OF IMPREGNATION IN POROUS MEDIA BY SELF-ORGANIZED PERCOLATION METHOD

## A.K. NGUYEN[1,2,4], E. BLOND[1], T. SAYET[1], A. BATAKIS[2], E. DE BILBAO[3], AND M.D. DUONG[4]

[1] Univ. Orléans, Univ. Tours, INSA CVL, LaMé – EA 7494
Polytech Orléans, 8 rue L. de Vinci, Orléans 45000, France
{anh-khoa.nguyen, eric.blond, thomas.sayet}@univ-orleans.fr

[2]Univ. Orléans, Institut Denis Poisson
UMR CNRS 6628, Av. De Blois, Orléans 45000, France
athanasios.batakis@univ-orleans.fr and http://www.univ-orleans.fr/mapmo/membres/batakis/

[3]CNRS, CEMHTI UPR 3079, Univ. Orléans
1D, av. de la Recherche Scientifique, F45071, Orléans Cedex 2, France
emmanuel.debilbao@univ-orleans.fr and http://www.cemhti.cnrs-orleans.fr

[4] University of Science, Ho Chi Minh City
227 Nguyen Van Cu, Vietnam
dmduc@hcmus.edu.vn and http://web.hcmus.edu.vn/en/index.php

**Abstract.** *The aim of this work is to develop a new numerical method to overcome the computational difficulties of numerical simulation of unsaturated impregnation in porous media. The numerical analysis by classical methods (F.E.M, theta-method, ...) for this phenomenon require small time-step and space discretization to ensure both convergence and accuracy. Yet this leads to a high computational cost. Moreover, a very small time-step can lead to spurious oscillations that impact the precision of the results. Thus, we propose to use a Self-organized Gradient Percolation (SGP) algorithm to reduce the computational cost and overcome these numerical drawbacks. The (SGP) method is based on gradient percolation theory, relevant to the calculation of local saturation. The initialization of this algorithm is driven by an analytic solution of the homogenous diffusion equation, which is a convolution between a Probability Density Function (PDF) and a smoothing function. Thus, we propose to reproduce the evolution of the capillary pressure profiles by the evolution of the standard deviation of the PDF. This algorithm is validated by comparing the results with the capillary pressure profiles and the mass gain curve obtained by finite element simulations and experimental measurements, respectively. The computational time of the proposed algorithm is lower than that of finite element models for quasi one-dimensional case. In conclusion, the SGP method permits to reduce the computational cost and does not produce spurious oscillations. The work is still going on for extension in 3D and the first results are promising.*

## 1 INTRODUCTION

The numerical modelling of the impregnation process requires a multiphysics model taking into account the material properties. Yet, it often demands large computing facilities.

The goal of this paper is to propose a new approach, which doesn't use classical modelling by partial differential equations and the associated numerical methods, to predict the capillary pressure profiles without spurious oscillations [1] and with reduced computational cost. To develop and present the basics of this method, the simplest case of non-reactive impregnation, for quasi one-dimensional problem, is developed.

A new numerical algorithm based on the gradient percolation theory is proposed. The initialization of the algorithm is driven by an analytic solution of the homogeneous diffusion equation, which is a convolution between a **P**robability **D**ensity **F**unction (PDF) and a smoothing function [2]. The evolution of the capillary pressure profiles with time is reproduced by the self-evolution of the standard deviation of the PDF. This model is therefore named **S**elf-organized **G**radient **P**ercolation model (SGP). In order to test this model, its solutions (i.e., the capillary pressure profiles and the mass gain curve) are compared with those obtained by F.E.M and by experiments, respectively.

## 2 SELF-ORGANIZED GRADIENT PERCOLATION (SGP) MODEL

### 2.1 Self-organized Gradient Percolation (SGP) model

The **S**elf-organized **G**radient **P**ercolation (SGP) model defines that the porous medium is considered as a random network model [3, 4] where each site $z$ has a local state $X(z)$ which should be transformed into the local average saturation and has Gaussian distribution [5] with mean $p(z)$ and variance $\theta^2$ as:

$$X(z) \sim \mathcal{N}(p(z), \theta^2) \tag{1}$$

The Eq. (1) shows that the local average saturation is driven by the PDF of Gaussian distribution that is used to construct the self-organization of the local profile such that:
(i) its initialization is proved as an analytic solution of the homogeneous diffusion equation [2];
(ii) for further time steps, the evolution of the capillary pressure profile should be reproduced by the self-evolution of the PDF.

### 2.2 Initialization of the SGP model

The SGP model aims at predicting the capillary pressure profile at any time that is inferred from a capillary pressure profile at initial time-step (i.e., initialization of the SGP model). As mathematically proven [2], the initialization is the analytic solution of the homogeneous diffusion equation (i.e. Richard's equation [6]) as the following formulation:

$$S(P_{cap}, t_0) = \mu(P_{cap}, t_0) * \delta(P_{cap}) \tag{2}$$

where $\langle * \rangle$ is convolution operator; $\delta(P_{cap})$ stands for a smoothing point-spread function; $\mu(P_{cap}, t_0)$ is the function of the local average saturation at initial time-step $t_0$, and is described as the PDF of a distribution with standard deviation $\sigma$, maximum saturation $S_{max}$ and residual saturation $S_r$ as:

$$\mu(P_{cap}, t_0) = S_r + (S_{max} - S_r) exp\left(-\frac{|P_{cap}|^m}{2\sigma^m}\right) \quad (3)$$

where $m$ characterizes the type of a distribution: $m = 1$ or $m = 2$ indicate that $\mu(P_{cap}, t_0)$ is the PDF for a Normal distribution or for a Laplace distribution by means of statistics, respectively.

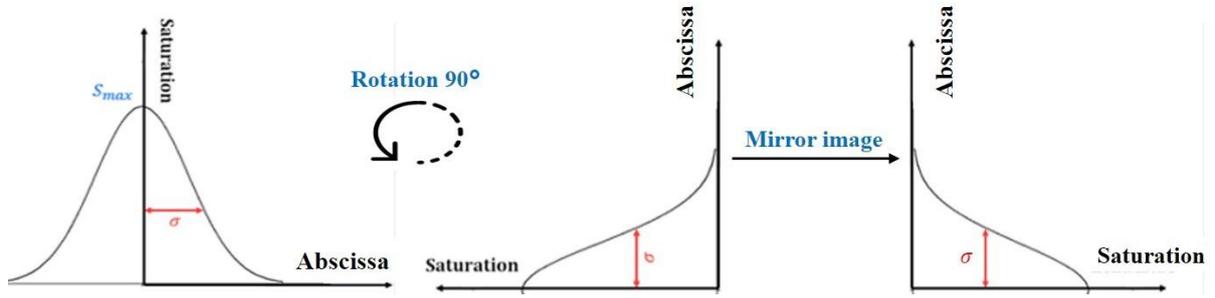

**Figure 1**: A simplified representation of the Eq. (3)

The Eq. (3) might be presumed to be more simple and convenient in application by the **Figure 1**. To determine the capillary pressure profiles, it is needed to identify the evolution of the local average saturation. It is thus proposed to reproduce the evolution of the local average saturation by that of the standard deviation of the PDF.

## 2.3 Time evolution

Let us remark that at each time-step $i$, a standard deviation function is denoted by $\sigma_i(P_{cap})$ to be different from constant standard deviation $\sigma_i$; the evolution of the abscissa $z$ with time-step $i$ is denoted by $\alpha(i, P_{cap})$. The evolution of standard deviation $\sigma_i$ is thus defined by:

$$\sigma_i = \sigma_{i-1} + \alpha(i, P_{cap}) \quad (4)$$

The evolution of the standard deviation function at the second time-step, i.e. $i = 1$, is expressed by the following form:

$$\sigma_1(P_{cap}) = \sigma_0 + [\sigma_1(P_{cap}) - \sigma_0] = \sigma_0 + (\sigma_1 - \sigma_0)f(P_{cap}) \quad (5)$$

where $f(P_{cap})$ is a function of the capillary pressure [7]. The general standard deviation function, i.e. $i = n$, is therefore deduced as follows:

$$\sigma_n(P_{cap}) = \sigma_0 + [\sigma_n(P_{cap}) - \sigma_0] = \sigma_0 + \left[\left(\sigma_{n-1} + \alpha(n, P_{cap})\right) - \sigma_0\right]f(P_{cap}) \quad (6)$$

At the local scale, the pore space can be assumed as a capillary (vertical for 1D model proposed herein). Thus, the evolution of the abscissa $z$ with time-step $n$, i.e. $\alpha(n, P_{cap})$, is driven by Poiseuille's equation [8] as follows:

$$\alpha(n, P_{cap}) = \frac{\partial z}{\partial n} = A \cdot P_{total}^n = A \cdot (P_{cap} + P_h^n) = A \cdot \Delta P \tag{7}$$

where $A$ depends on capillary diameter and liquid properties, $P_{total}^n$ and $P_h^n$ are the total and the hydrostatic pressure at time-step $n$, respectively; $\Delta P$ is the difference between capillary and hydrostatic pressures. In the case of the vertical capillary tube, the Eq. (7) points out that the transient simulation stops at the steady-state when there is no longer difference between capillary pressure and hydrostatic pressure (i.e., $\Delta P = 0$).

## 2.4 Continuity and Boundary Conditions

As shown in the Eq. (2), a convolution procedure is first employed to introduce the continuity of the SGP model. Indeed, the procedure is related to the local average saturation and the smoothing point-spread function as the following formulation:

$$S(z) = (\mu * \delta)(z) = \sum_{z' \in \Omega} \mu(z - z')\delta(z') \tag{8}$$

where $\delta(z') = \frac{1}{N}\sum \mathcal{X}(z')$, $\Omega = \{z' \in \mathbb{Z}^2: |z'| \leq 1\}$, $\mathcal{X}(z')$ is the Kronecker function, and $N$ is number of site $z \in \Omega$.

For impregnation in porous media, several boundary conditions related to the physical exchange made on the concerned interface are distinguished by: the case where liquid can/ cannot flow out of the boundary (i.e., the constant air pressure and the draining condition/ the undrained condition, respectively). The choice of the boundary conditions depends on the interpretation of the physical phenomena. In the SGP model, the convolution of the values on the boundary surface with the smoothing function is then employed to model the boundary conditions.

## 3 THE VALIDATION OF THE SGP METHOD

One-dimensional capillary rising tests were performed on porous samples [9]. The amount of impregnated liquid into the porous medium was measured by weighing the loss of liquid in the bath as can be seen in **Figure 2**. Porous material and liquid for the tests are reported in **Table 1**.

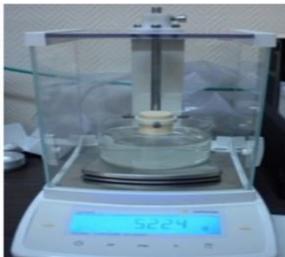

**Figure 2**: The non-reactive impregnation test

**Table 1**: Materials and liquid for the tests

| Test | Porous sample | Liquid |
|------|---------------|--------|
| 1 | Alumina 99% | Glycerine |
| 2 | Alumina 99% | Oil |

To validate the SGP method, a **F**inite **E**lement (F.E.) model has been implemented in the software "ASTER" [9].

**Table 2**: Values of physical chracteristics used of the porous materials and the liquid

| Properties | Values | | Units |
|---|---|---|---|
| | Test 1 | Test 2 | |
| Initial porosity | 0.2 | 0.19 | |
| Viscosity of the fluid, $\eta$ | 1.02 | 0.35 | $Pa \cdot s$ |
| Mass density of fluid, $\rho_W$ | 1260 | 892 | $kg \cdot m^{-3}$ |
| Intrinsic permeability, $K_{int}$ | $9.5 \times 10^{-12}$ | $7.44 \times 10^{-11}$ | $m^2$ |

To fit the capillary pressure curve, two well-known models (c.f. Brook's model and van Genuchten's model) [10, 11] are employed where pressure reference $P_0$ or $P_e$ and empirical parameter $l$ or $\lambda$ need to be determined. To apply the algorithm of the SGP model on both tests, the values of input data are reported in **Table 3**.

**Table 3**: Values of input data used for the SGP algorithm

| Test | $A\ (m/s)$ | $l$ or $\lambda$ | $P_0$ or $P_e$ (Pa) |
|---|---|---|---|
| 1 | $1,16 \cdot 10^{-7}$ | 0,55 | 1100 |
| 2 | $1,9 \cdot 10^{-6}$ | 2,62 | 4870 |

## 3.1 Comparisons between numerical results and experimental data

For the tests 1 and 2, because there is no experimental data for the capillary pressure profiles, the mass gain curve obtained through the SGP model is compared with that obtained experimentally [9] (**Figure 3**).

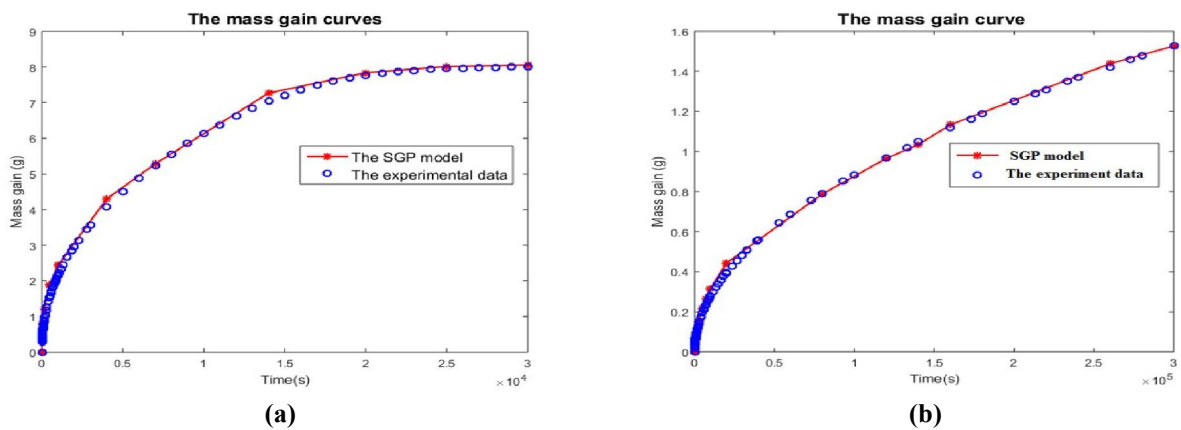

**Figure 3**: Measured and simulated mass gain curve: **(a)** for the test 1 and **(b)** for the test 2.

The **Figure 3** shows that numerical results from the SGP algorithm fit well with experimental data.

### 3.2 Comparisons between F.E and SGP results

For the tests 1 and 2, the comparisons of the capillary pressure profiles between the SGP model and the FEM model are shown in **Figure 4**.

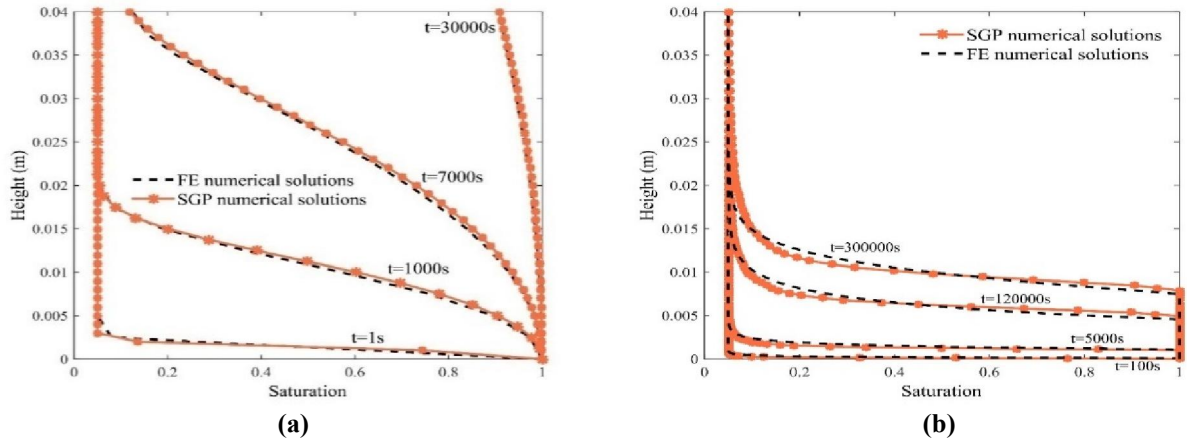

**(a)** **(b)**
**Figure 4**: The evolution of the simulated capillary pressure profiles: **(a)** for the test 1 and **(b)** for the test 2.

The **Figure 4** shows that F.E. results fit well with SGP results too. Concerning the comparisons between F.E. and SGP results for the test 2, the CPC from Brook's model is employed, which is assumed to be a PDF of the Laplace distribution, i.e. $m = 1$ for the SGP model. Thus, the capillary pressure profiles obtained by the SGP model always remain the form of this PDF. That's why the F.E. and SGP solutions are not exactly the same (as can be seen in the **Figure 4 (b)**).

The computational costs for the two tests are also reported in **Table 4.** By comparisons of the CPU time, the **Table 4** shows that the computational time of the SGP model is lower than the one of F.E.M.

**Table 4 :** Computational time of the SGP model and the F.E model.

| Test | CPU time (seconds) | |
|---|---|---|
| | **SGP model** | **F.E model** |
| **1** | 6.52 | 25.5 |
| **2** | 1.1 | 12.1 |

Finally, the objectives of this study are entirely satisfied: (i) the solutions of the SGP method are free from spurious oscillations and (ii) the computational time is reduced.

### 4 CONCLUSIONS

In this study, a novel algorithm based on the gradient percolation theory, named Self-organized Gradient Percolation (SGP) algorithm, is proposed. Here, the gradient percolation

A.K. Nguyen, E. Blond, T. Sayet, A. Batakis, E. de Bilbao and M.D. Duongtheory is used to compute the local saturation. The link between the mathematics and the physical issue (i.e., the non-reactive impregnation phenomenon) is done thanks to the initialization of the algorithm by an analytic solution of the Richard's equation. Furthermore, the convolution operator allowed to ensure the spatial continuity of the wetting fluid and to check the boundary conditions of the problem. The comparisons with experimental measurements show a good agreement for the quasi one-dimensional case of vertical impregnation. The comparisons with F.E.M exhibits a lower computational cost for the SGP method. First results are promising with respect to possible generalization to 3D case.

**REFERENCES**

[1] P. A. Verruijt and A. Vermeer, "An accuracy condition for consolidation by finite elements," *International Journal for Numerical and Analytical Method in Geomechanics,* vol. 5, no. 1, pp. 1-14, 1981.

[2] B. Jähne, H. Haubecker and G. P., Handbook of Computer Vision and Applications, San Diego : Academic Press, 2000.

[3] G. Grimmett, Percolation, Springer, 1999.

[4] P. Nolin, "Critical exponents of planar gradient percolation," *The Annals of Probability,* vol. 36, no. 5, pp. 1748-1776, 2008.

[5] N. Nguyen and A. Batakis, "Simulations of the growth of cities," PhD thesis of University of Orleans, Orleans, 2014.

[6] M. A. Celia, E. T. Bouloutas and R. L. Zarba, "A general mass-conservative numerical solution for the unsaturated flows equation," *Water Resources Research,* vol. 26, no. 7, pp. 1483-1496, 1990.

[7] R. Christiansen, J.-S. Kalbus and S.-M. Howarth, "Evaluation of methods for measuring relative permeability of Anhydrite from the Salado Formation: Sensitivity Analysis and Data Reduction," Sandia national Laboratoires, California, 1997.

[8] C. W. Extrand, "Forces, pressures and energies associated with liquid rising in nonuniform capillary tubes," *Journal of Colloid and Interface Science,* vol. 450, pp. 135-140, 2015.

[9] E. de Bilbao, Y. Hbiriq, C. Orgeur, S. Brassamin, J. Poirier, L. Loison and T. Thonnesen, "Identification of Transport Properties of Refractories: Intrinsic permeability and Capillary Pressure Curve," in *59th Colliquium on refractories*, Aechan (Germany), 2016.

[10] R. Brooks and A. Corey, "Hydraulic Properties of Porous Media," Colodaro State University, Fort Collins, Colodaro, 1964.

[11] M. T. van Genuchten, "A closed-form for predicting the hydraulic conductivity of unsaturated soils," *Soil Science Society of America,* vol. 44, no. 5, pp. 892-898, 1980.7